\documentclass[aps,prl,twocolumn,superscriptaddress,showpacs,showkeys,floatfix]{revtex4-1}
\usepackage{graphicx}
\usepackage{amssymb}
\usepackage{amsmath}
\usepackage{bm}
\begin{document}
\setcounter{page}{1}
\title{Interspecific competition underlying mutualistic networks}
\author{Seong Eun \surname{Maeng}}
\author{Jae Woo \surname{Lee}}
\email{jaewlee@inha.ac.kr}
\affiliation{Department of Physics, Inha University, Incheon 402-751, Korea}
\author{Deok-Sun \surname{Lee}}
\email{deoksun.lee@inha.ac.kr}
\affiliation{Department of Physics, Inha University, Incheon 402-751, Korea}
\affiliation{Department of Natural Medical Sciences, Inha University, Incheon 402-751, Korea}

\begin{abstract}
Multiple classes of interactions may exist affecting one another in a given system.
For the mutualistic networks of plants and pollinating animals, it has been known that the degree distribution is 
broad but often deviate from power-law form more significantly for plants than animals. 
To illuminate the origin of such asymmetry,
we study a model network in which links are assigned under generalized preferential-selection rules between two 
groups of nodes and find the sensitive dependence of the resulting connectivity pattern on the model parameters. 
The nonlinearity of preferential selection can come from interspecific interactions among animals and among plants. 
The model-based analysis of real-world mutualistic networks  suggests that  a new animal determines its partners not only by their abundance but also under the competition with existing animal species, which leads to the stretched-exponential degree distributions of plants.
\end{abstract}


\maketitle 

Diverse interactions and dependencies among nonidentical elements are characteristic of complex systems~\cite{mucha10, buldyrev10,ahn10}.  
Ecological systems are a prototypical example, in which numerous species interact via predation, herbivory, 
mutualistic support, competition, cooperation, and so on, and their network structure and function have 
attracted much attention~\cite{pascual05,bascompte10,montoya06,saavedra09,bastolla09}. 
The mutualistic relation between two species is beneficial for the survival and reproduction of both of them 
such as animal pollinators and flowering plants. 
Given numerous species of  plants and insects, producing nectar with different composition and flavor and 
carrying a wide range of preference and capacities, respectively, the establishment of individual mutualistic relationship 
depends on specific needs and qualification. Nevertheless, the global organization of  the mutualistic plant-pollinator networks exhibits 
common features~\cite{jordano03,bascompte03,vazquez05,guimaraes07,santamaria07,guimaraes07,maeng11,campbell11}. 
The degrees of plant and animal species are  distributed broadly in general,  but their distributions often deviate from 
a power-law form, more significantly for plant species: While the degree distribution of animals are close to power laws, 
those of plants are of  truncated power-law, exponential, or stretched-exponential form~\cite{jordano03,guimaraes07,maeng11}. 
Biological matching, species abundance, and the difference between the numbers of animals and plants may bring such deviation 
from scale invariance~\cite{jordano03,vazquez05,guimaraes07}, which, however, remains to be addressed further~\cite{santamaria07}. Interestingly, mutualistic networks have their topological features distinguished from 
trophic networks particularly in their nestedness and modularity, which is related to the stable architecture varying 
with the type of interaction~\cite{thebault10}.  

To gain insight into the underlying mechanism of mutualistic community formation, 
here we propose and study a simple growing bipartite network model and apply it to analyze real-world mutualistic networks. 
The model is based on a generalized preferential-selection rule, being an extension of the model in Ref.~\cite{guimaraes07}.  
From the nestedness and broad degree distributions identified in lots of mutualistic communities, the preferential 
selection~\cite{barabasi99} has been expected to play a role in their evolution~\cite{jordano03,guimaraes07,santamaria07,thebault10}. 
Our model provides a unified picture of evolving mutualistic communities. The nonlinear preferential selection 
reflects the impact of individual characteristics and interspecific interactions on determining symbionts. 
Fitting the model prediction to empirical data, we discover a pattern of interspecific interaction 
affecting the architecture of mutualistic networks. While a new animal species is attracted to a plant species with high abundance and thus with many pollinators already, it should compete with the existing pollinators. 
As a result, the chance to find a plant species with a large number of pollinators is not as high as expected, which is shown to cause the degree distribution of plants to take a stretched-exponential form.

\begin{figure}
\includegraphics[width=6cm]{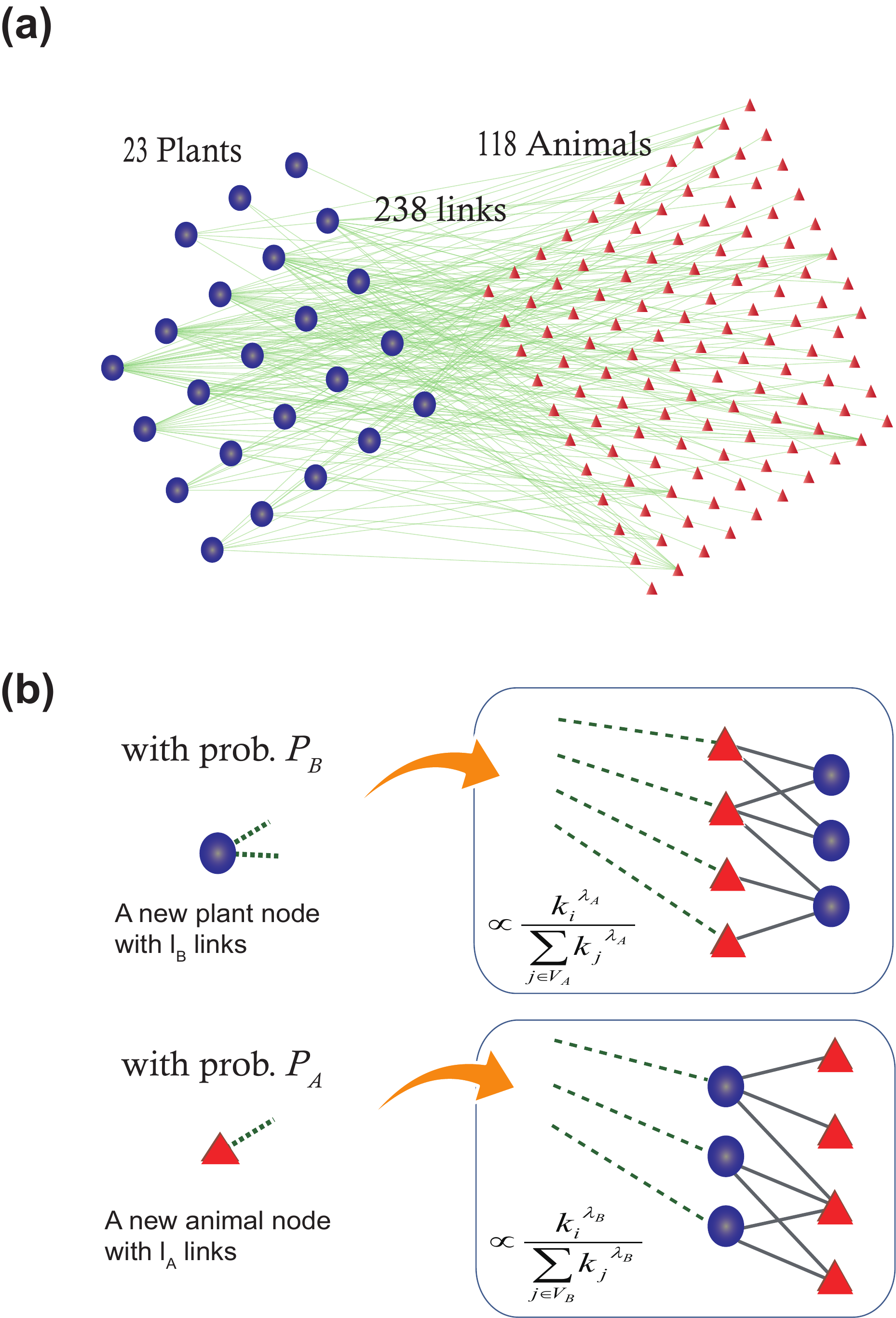}
\caption{Plant-pollinator mutualistic network and its model. (a) A mutualistic network at a subarctic-alpine site in northern Sweden~\cite{elberling99, iwdb}. (b) GBN model. A new animal (plant) node, introduced with probability $P_A \  (P_B)$ every time step, selects $\ell_A \ (\ell_B)$ plant (animal) nodes one by one, each with probability proportional to the $\lambda_B \ (\lambda_A)$ th power of its degree.}
\label{fig:model}
\end{figure}

To be specific, we consider plant-pollinator mutualistic networks as in Fig.~\ref{fig:model} (a),  in which each node represents a species of either animal ($A$) type or plant ($B$) type and some pairs of nodes of the opposite types are connected representing their mutualistic relationship. 
A growing bipartite network (GBN) model defined below and also sketched in Fig.~\ref{fig:model}(b)~\cite{guimaraes07,ramasco04,goldstein05,lambiotte05} 
illustrates the topological evolution of the plant-pollinator networks.
Initially  there are $\ell_A$ nodes of type $A$ and $\ell_B$ nodes of type $B$,  all pairs of nodes of the opposite types being connected. At each time step, a new node of type $A \ (B)$ arrives with probability $P_A \  (P_B)$, where $P_A + P_B = 1$. The new node of type $A$ ($B$) are linked to $ \ell_{A} \ (\ell_B) $ partners of type $B\ (A)$ that are selected with the probability proportional to the $\lambda_B \ (\lambda_A)$ th power of their degrees.  Iterating these procedures up to time $N$,  one obtains a bipartite network of $NP_A$ nodes of type $A$ and $NP_B$ nodes of type $B$ on average.

%%%%%%%2nd Revision%%%%%
The degree-based preferential selection  by a new species is assumed in this model and can be motivated as follows.
In case that  a new species appears by speciation from an existing one, it can be assumed to inherit the mutualistic interactions of the ancestor species, as the protein interactions are inherited by duplicated genes~\cite{berg04}. Then, the more partners an existing species has, the higher the chance to have a new mutualistic partner speciated from one of its old partners is. 
Also, if we assume that a new species in a community,  appearing by either speciation or migration, selects randomly its partner organisms,  the new species will be more likely to form a mutualistic relationship with a more abundant species due to the existence of more organisms of the species. Given that the number of mutualistic partners - degree - of a species is  positively correlated with the abundance of the species, the degree-based preferential selection is expected to work in this case, too. 
% which should have larger degrees on average. 
%%%%%%%%%%%%%%%%%%%%%%%%%%%%%%%%%%%%%

Restricted resources provided by one type of species may draw competition among the other type of species, which can give rise to the nonlinear preferential selection.   For a new animal species $a^{\rm (new)}$, let us assign $r^{(b)}_a$ to each pollinator $a$ of each plant $b$, which  can be negative or positive depending on the significance of the competitive interaction between $a$ and $a^{\rm (new)}$ relative to the abundance of plant $b$.
The sum of those interaction  coefficients over $a$ can be a measure of the chance that $a^{\rm (new)}$ can occupy the plant $b$. When the sum is negative, the plant will not be selected. Thus, the selection probability of plant $b$ with degree $k$ will be proportional to $K_b = \max\{0,\sum_{a=1}^k r^{(b)}_a\}$, which can be viewed as the effective degree of $b$ for $a^{\rm (new)}$. On ensemble average, $K_b$ and the selection probability will be proportional to $k$ if $\langle r_a^{(b)}\rangle>0$. On the other hand, if $\langle r^{(b)}_a\rangle=0$, $\langle K_b\rangle \sim k^{1/d_w}$, in which $d_w$ is the random-walk dimension varying with the distribution and correlation of $r_a^{(b)}$'s~\cite{hughesbook}. Both cases are considered in our model with $0\leq \lambda_B\leq 1$. Similarly, a new plant $b^{\rm (new)}$ can be complementary or substitutable for each pollinator $b$ nurturing each animal $a$, which leads the probability of animal $a$ to pollinate $b^{\rm (new)}$ to scale with $k$ linearly or sublinearly. The impact of such nonlinear preferential attachment has been investigated for unipartite growing networks~\cite{krapivsky00} and the preferential-selection exponent, $\lambda$, has been measured empirically e.g., for the Internet~\cite{satorras01}. 

\begin{figure}
\includegraphics[width=\columnwidth]{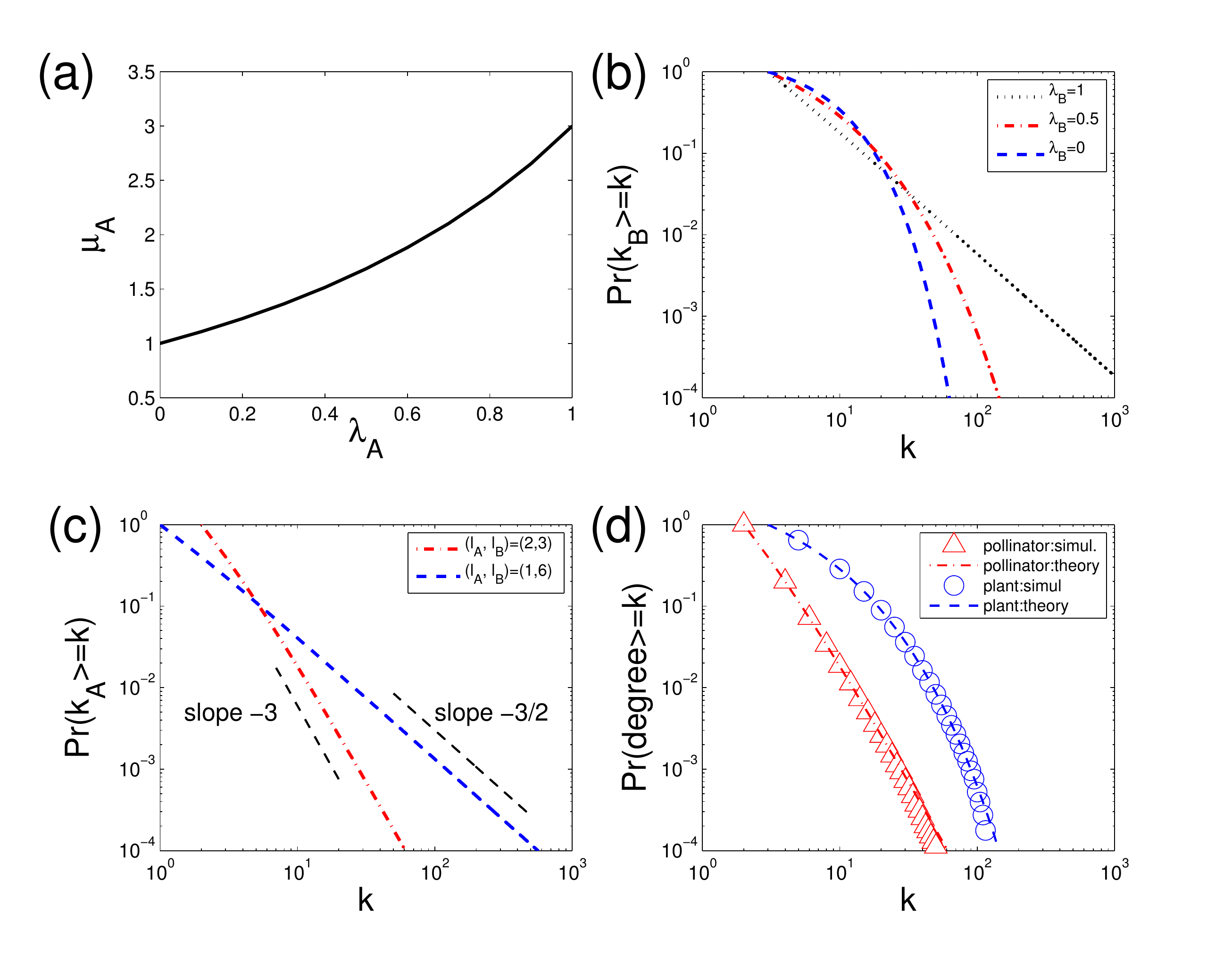}
\caption{Degree distributions  in the GBN model.
(a) Plot of $\mu_A$ versus $\lambda_A$ with $(\ell_A,\ell_B)=(2,3)$. 
(b) Cumulative degree distributions ${\rm Pr} (k_B\geq k) = \sum_{k_B\geq k} n_B(k)$ of plant nodes for different values of $\lambda_B$.  $(\ell_A,\ell_B)=(2,3)$ for all cases.
(c) Cumulative degree distributions of animal nodes for  $(\ell_A,\ell_B) = (2,3)$ and $(1,6)$ take power laws $k^{-3}$ and $k^{-3/2}$ asymptotically. $\lambda_A=1$ for both cases. 
(d) Comparison of simulations (points) and  theory (lines) from Eqs.~(\ref{eq:nk}) and (\ref{eq:mu}). Here $\ell_A = 2, \ell_B=3, \lambda_A=1$, and $\lambda_B=0.5$.  For simulations, we averaged the cumulative degree distributions over  $1000$ networks of $N= 10^4$ nodes. In (a-d), we set $P_A = 3/4$ and $P_B = 1/4$.}
\label{fig:simul}
\end{figure}

The mean number $N_{X}(k,\tau)$ of nodes of degree $k$ and type $X$ evolves with time as~\cite{krapivsky00} 
\begin{equation}
\begin{split}
& N_X (k,\tau +1) - N_X (k,\tau )=  P_X \delta_{k,\ell_X} +\\ 
& {P_Y \ell_Y \over M_X(\tau)}\left[ (k-1)^{\lambda_X} N_X(k-1,\tau)-k^{\lambda_X} N_X(k,\tau)\right],
\end{split}
\label{eq:Nkt}
\end{equation}
where $M_{X}(k,\tau) =  \sum_{j\in V_X} k_j^{\lambda_X} = \sum_{k=\ell_X} k^{\lambda_X} N_X(k,\tau)$. We use  $X$ and $Y$ to indicate the opposite types of nodes such that $(X,Y)=(A,B)$ or $(B,A)$.  In the long-time limit $\tau\gg 1$, 
$N_X(k,\tau)$ is proportional to time $\tau$ such that $N_X(k,\tau)= \tau P_X n_X(k)$, where $n_X(k)$ is the degree distribution for type $X$. It is inserted into Eq.~(\ref{eq:Nkt}) to yield   
\begin{equation}
n_X(k) = {\mu_X \over k^{\lambda_X}} \prod_{q = \ell_X}^k {q^{\lambda_X}\over q^{\lambda_X} + \mu_X} \ {\rm for}\ k\geq \ell_X,
\label{eq:nk}
\end{equation}
where the constant $\mu_X$ is given by  
\begin{equation}
\mu_X = {P_X\over P_Y \ell_Y}\sum_{k=\ell_X}^\infty k^{\lambda_X} n_X(k).
\label{eq:mu}
\end{equation}
It is worthy to note that Eq.~(\ref{eq:Nkt}) has the Kramers-Moyal coefficients that are identical at all orders~\cite{riskenbook}, preventing a truncated expansion. While $\lambda_A$ and $\lambda_B$ are engaged in determining $\mu_A$ and $\mu_B$, respectively, both $\mu_A$ and $\mu_B$ depend on $P_A (=1-P_B), \ell_A$, and $\ell_B$, leading $n_A(k)$ and $n_B(k)$ to be coupled. Note that $\mu_X =  P_X/(P_Y \ell_Y)$ at $\lambda_X=0$ and $\mu_X = 1 + P_{X}\ell_{X}/(P_{Y}\ell_{Y})$ at $\lambda_X=1$ (Fig.~\ref{fig:simul} (a)).

The dependence of the degree distributions on the parameters is manifested in its asymptotic behavior: 
\begin{eqnarray}
\log n_X(k) &=&  \ln \left({{\mu}_X\over k^{\lambda_X}}\right) - \sum_{q=\ell_X}^k \ln\left(1+ {\mu_X\over q^{\lambda_X}}\right) \nonumber\\
&\simeq& \ln \left({{\mu}_X\over k^{\lambda_X}}\right) - \sum_{q=\ell_X}^k {\mu_X \over q^{\lambda_X }} \ \ (k\gg 1)\nonumber\\
&\sim& \left\{
\begin{array}{ll}
-\gamma_X \ln k  & {\rm for} \ \lambda_X = 1,\\
 -\mu_X {k^{1-\lambda_X} \over 1-\lambda_X} & {\rm for} \ 0\leq \lambda_X<1,
\end{array}
\right.
\label{eq:asym}
\end{eqnarray}
with 
\begin{equation}
\gamma_X = 1+ \mu_X = 2 + {P_X \ell_X\over P_Y \ell_Y}. 
\label{eq:degexp}
\end{equation}
Remarkably, $n_X(k)$ changes from a power-law to a stretched-exponential form as $\lambda_X$ deviates from $1$ as in the example in Fig.~\ref{fig:simul} (b)~\cite{krapivsky00}. 
For $\lambda_X=1$, the degree exponent $\gamma_X$ of power-law degree distributions varies continuously not only with the fraction of nodes of the two types $P_A$ and $P_B$~\cite{guimaraes07} but also with the initial numbers of links $\ell_A$ and $\ell_B$ (Fig.~\ref{fig:simul}(c)).  Such sensitive dependence in bipartite networks enables us to look into the microscopic details through the global organization of bipartite systems. 
The numerical solutions are in agreement with model simulations  as in Fig.~\ref{fig:simul}(d). 

If $\ell_A$ and $\ell_B$ are integers, the possible value of  $\gamma_X$ in  Eq.~(\ref{eq:degexp}) are restricted, which is not the case in real-world networks. Considering $\ell_X$ as the mean number of initial links, we let a new node of type $X$ assigned $\lfloor \ell_X\rfloor$ or $\lfloor \ell_X\rfloor +1$ links with probability  $\lfloor \ell_{X}\rfloor +1-\ell_{X}$ or $\ell_{X}-\lfloor \ell_{X}\rfloor$, respectively.  $\lfloor x\rfloor$ is the largest integer not larger than $x$. Then  Eq.~(\ref{eq:asym}) remains to be true, with a slight change of $n_X(k)$ from Eq.~(\ref{eq:nk}) to
\begin{equation}
\begin{split}
& n_X (k) =\\ 
& \left\{
\begin{array}{ll}
{\mu_X \over k^{\gamma_X}} \left[1+\mu_X {\ell_X -\lfloor\ell_X\rfloor \over \lfloor\ell_X\rfloor^{\gamma_X}}\right]\prod_{q=\lfloor\ell_X\rfloor}^k {q^{\lambda_X}\over q^{\lambda_X} + \mu_X} & (\ k> \lfloor\ell_X\rfloor ), \\
{\mu_X (\lfloor\ell_X\rfloor +1 -\ell_X)\over \lfloor\ell_X\rfloor^{\lambda_X} + \mu_X} & ( k = \lfloor\ell_X\rfloor)  
\end{array}
\right.
\end{split}
\label{eq:nk2}
\end{equation}
with $\mu_X = [P_X/(P_Y \ell_Y)]\sum_{k=\lfloor\ell_X\rfloor}^\infty k^{\lambda_X} n_X(k)$.

\begin{figure*}
\includegraphics[width=2\columnwidth]{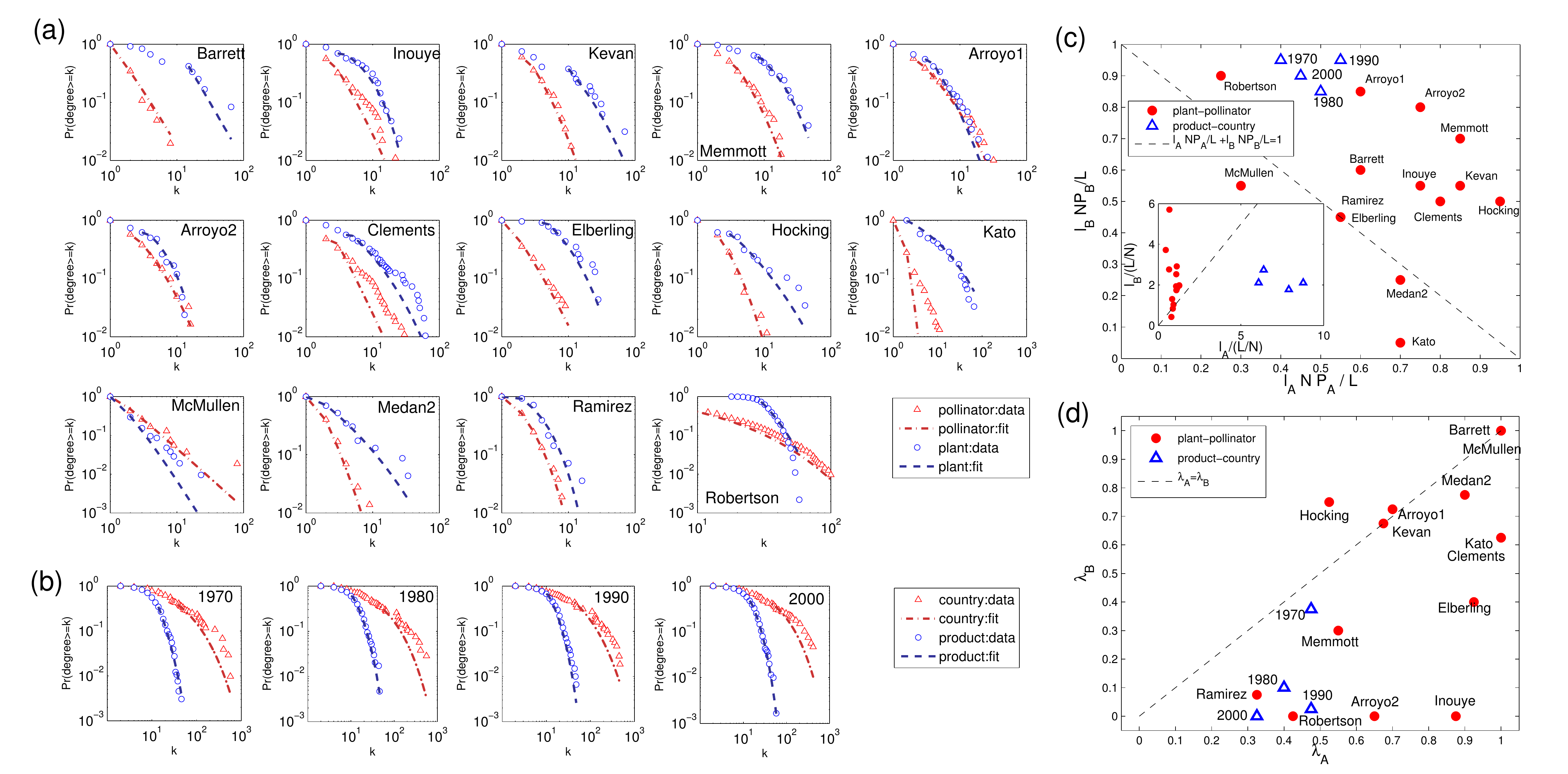}
\caption{Fitting model degree distributions in Eq.~(\ref{eq:nk2}) to real-world networks. (a) Degree distributions of plant-pollinator networks from different communities (points), denoted by  ID made after the first author of each reference~\cite{iwdb}, and their maximum-likelihood fits (lines) of the GBN model. There are 81 to 1881 nodes in these networks. (b) Degree distributions of the product-country networks of 1621 products and 132 countries in different years. If the export volume of a product by a country is in the top $25\%$ among all the export volumes, they are connected. (c) Normalized initial numbers of links. They are rather uniformly distributed mostly in the region $\ell_A NP_A/L + \ell_BNP_B/L \geq 1$. (inset) For plant-pollinator networks, $\ell_A<\ell_B$ except for two networks, ``Kato'' and ``McMullen''. For product-country networks, $\ell_A>\ell_B$. (d) Preferential-selection exponents. $\lambda_B<\lambda_A$ in most networks. For Hocking plant-pollinator network, the competition between flowers is known to be more significant than between pollinators~\cite{hocking68}, leading to $\lambda_A<\lambda_B$ in our model.  }
\label{fig:analysis}
\end{figure*}

We fitted the model  predictions of  the degree distributions in Eq.~(\ref{eq:nk2}) to those of $14$ real plant-pollinator networks~\cite{iwdb} by adjusting $\ell_A, \ell_B, \lambda_A$, and $\lambda_B$ through the combined use of the maximum-likelihood estimation and the Kolmogorov-Smirnov statistic~\cite{clauset09}. 
The fitted degree distributions are in good agreement with data as shown in Fig.~\ref{fig:analysis} (a). The initial number of links of plants $\ell_B$ turns out to be much larger than that of animals $\ell_A$  probably due to 
 there being more animal than plant species; their normalized values, $\ell_A N P_A/L$ and $\ell_BNP_B/L$, are distributed uniformly as shown in Fig.~\ref{fig:analysis} (c).  Most importantly, the preferential-selection exponents show a significant asymmetry: $\lambda_B < \lambda_A$ in most  networks as shown in Fig.~\ref{fig:analysis} (d). 
This implies the significantly strong competition between a new animal and existing pollinators contrary to the relatively weak competition between plants. Thus one can see that the restriction on the number of available plant species is a more crucial factor in shaping the mutualistic community than the restriction on available animal species, possibly related to the difference in their survival and reproduction rates. Consequently, plants with large degrees have the advantage of their high abundance screened by the competition between animals characterized by $\lambda_B$ less than $1$, which 
leads the degree distribution to take a stretched-exponential form $n_B(k)\sim e^{-{\rm (const.)} k^{1-\lambda_B}}$. We remark that these observations are not the case for ``Hocking'' in Fig.~\ref{fig:analysis} (d), for which it has been reported in Ref.~\cite{hocking68} that the competition between plants is more significant than between pollinators, implying $\lambda_A<\lambda_B$ . 

 To check further the robustness of our results, we analyzed  the product-country networks representing which country exports which product~\cite{hidalgo07}. Both the prevalence of a product and the wealth of each country depend on trade volume  and thus a country (A) and a product (B) may have mutualistic relation in the global economic ecosystems. Although $P_A<P_B$ and $\ell_A> \ell_B$  contrasted to plant-pollinator networks, we find again that $\lambda_B<\lambda_A$ suggesting the same patterns of competition among countries as among pollinators.

In conclusion, we found that the connectivity pattern of bipartite networks is sensitively dependent on model details. Such sensitivity can be of potential use in controlling transport and epidemic affected by the connectivity in multiconnected systems and, in the present work, allowed us to illuminate the interaction patterns hidden under the bipartite architecture of mutualistic networks. The stretched-exponential form of the degree distribution of plants is shown to be driven by the significant competition between pollinating animals, which is true also for countries in economic mutualistic networks. 
Our results demonstrate the importance of taking into account the cross-talk between different types of links interwoven in a system in order to understand   and predict its properties. While we focused on the global network architecture of mutualistic communities, it is highly desirable to extend our model to account for the population dynamics  and diverse interspecific interactions.

\begin{acknowledgments}
We thank the anonymous reviewers for helpful comments. This work was supported by 
Basic Science Research Program through the National Research Foundation of Korea (NRF)
funded by the Ministry of Education, Science and Technology [Grants No. 2011-0003015 (J.W.L) and
No. 2011-0003488 (D.-S.L)]. D.-S.L. acknowledges the TJ Park Foundation for support. 
\end{acknowledgments}

\end{document}